\documentclass[aps,pra,twocolumn,groupedaddress,showpacs]{revtex4}

\usepackage{amssymb}
\usepackage{graphicx}
\usepackage[dvips]{epsfig}
\usepackage{bm}
\usepackage{amsmath}
\usepackage{wrapfig}

\usepackage{enumitem}

\usepackage{color}
\begin{document}
\title{Demonstration of a state-insensitive, compensated nanofiber trap}

\author{A. Goban$^{1\ast}$, K. S. Choi$^{1,2\ast}$, D. J. Alton$^{1\ast}$, D. Ding$^1$, C. Lacro\^ute$^1$, M. Pototschnig$^{1}$, T. Thiele$^{1\dagger}$, N. P. Stern$^{1\ddag}$, and H. J. Kimble$^1$}
\address{$^1$ Norman Bridge Laboratory of Physics 12-33, California Institute of Technology, Pasadena, California 91125, USA}
\address{$^2$ Spin Convergence Research Center 39-1, Korea Institute of Science and Technology, Seoul 136-791, Korea}
%\date{\today}
%\pacs{42.50.Ct, 37.10.Gh, 37.10.Jk, 42.50.Ex}

\begin{abstract}
We report the experimental realization of an optical trap that localizes single Cs atoms $\simeq 215$ nm from surface of a dielectric nanofiber. By operating at magic wavelengths for pairs of counter-propagating red- and blue-detuned trapping beams, differential scalar light shifts are eliminated, and vector shifts are suppressed by $\approx250$. We thereby measure an absorption linewidth $\Gamma/2\pi = 5.7 \pm 0.1$ MHz for the Cs $6S_{1/2},F=4$ $\rightarrow$ $6P_{3/2},F'=5$ transition, where $\Gamma_0/2\pi = 5.2$ MHz in free space. Optical depth $d \simeq 66$ is observed, corresponding to an optical depth per atom $d_1\simeq 0.08$. These advances provide an important capability for the implementation of functional quantum optical networks and precision atomic spectroscopy near dielectric surfaces.
\end{abstract}

%\date{\today}
\maketitle

An exciting frontier in quantum information science is the integration of otherwise ``simple'' quantum elements into complex quantum networks \cite{kimble08a}. The laboratory realization of even small quantum networks enables the exploration of physical systems that have not heretofore existed in the natural world. Within this context, there is active research to achieve lithographic quantum optical circuits, for which atoms are trapped near micro- and nano-scopic dielectric structures and ``wired'' together by photons propagating through the circuit elements. Single atoms and atomic ensembles endow quantum functionality for otherwise linear optical circuits and thereby the capability to build quantum networks component by component.

Creating optical traps compatible with the modal geometries of micro- and nano-scopic optical resonators and waveguides \cite{vahala-review,eichenfield_optomechanical_2009} is a long-standing challenge in AMO physics \cite{balykin91,vernooy97,burke_designing_2002}. `Optical tweezers' with tight focussing have succeeded in trapping single atoms within small volumes $\sim \lambda^3$  \cite{schlosser_sub-poissonian_2001}, but the focal geometries of conventional optical elements are not compatible with atomic localization $\simeq 100$ nm near microscopic photonic structures \cite{vahala-review,eichenfield_optomechanical_2009}. Moreover, spatially inhomogeneous energy shifts $U(\mathbf{r})$ on a sub-wavelength scale generally depend on the atomic internal state, limiting long-lived trap and coherence times \cite{Corwin1999}. Nevertheless, important advances have been made by loading ultracold atoms into hollow-core optical fibers  \cite{Renn1995,Ito1996,Christensen2008,londero09,Bajcsy2009} and by trapping atoms in the evanescent fields of nanoscale waveguides \cite{Balykin2004,Kien2004a,Kien2005a,Nayak2007,Sague2007,Vetsch2010,Vetsch2010a,Dawkins2011,Chang2009}.

\begin{figure}[t!]
%\vspace{-5mm}
\centering
\includegraphics[width=1.0\columnwidth]{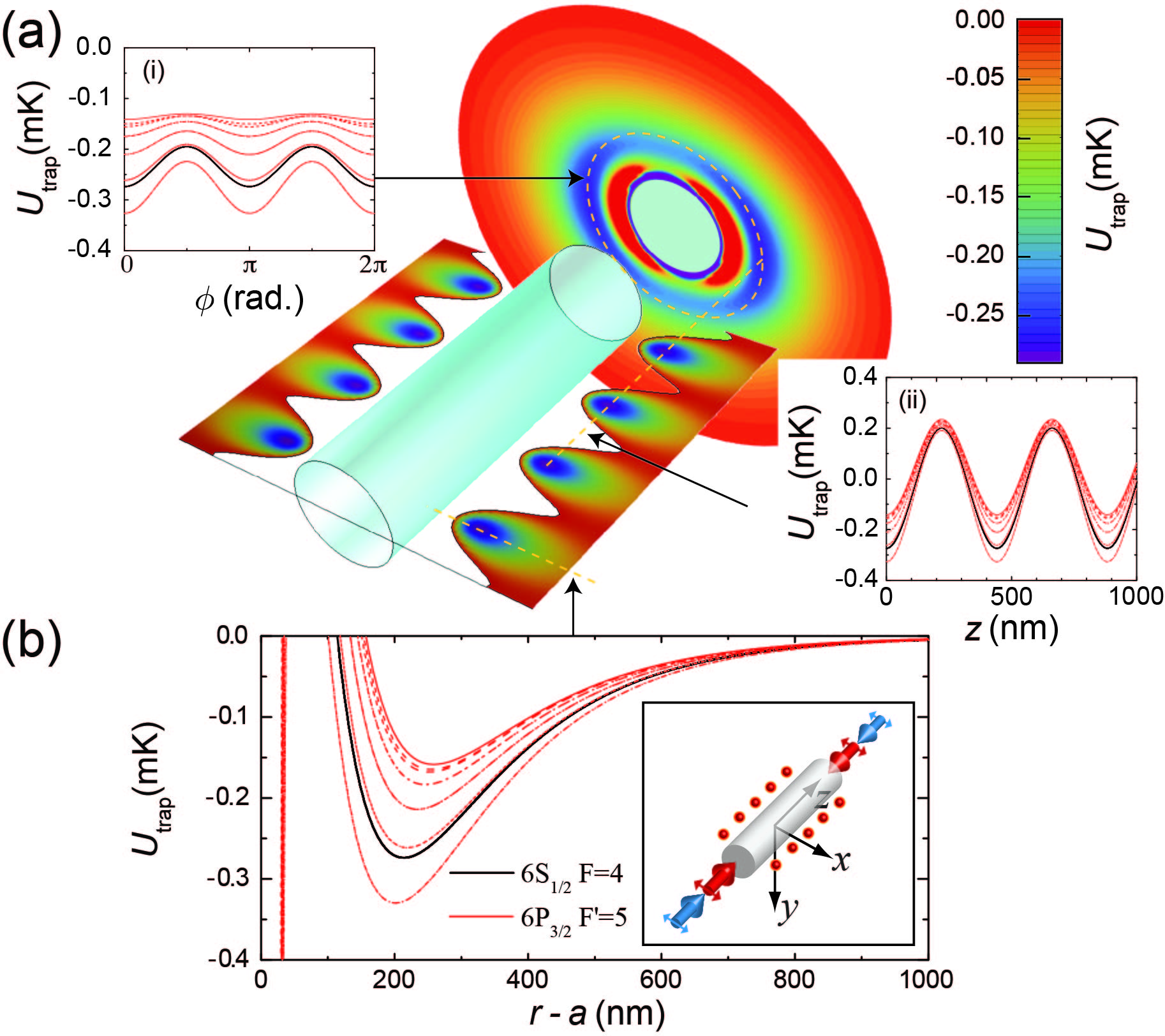}
\caption{Adiabatic trapping potential $U_{\text{trap}}$ for a state-insensitive, compensated nanofiber trap for the $6S_{1/2}, F=4$ states in atomic Cs outside of a cylindrical waveguide of radius $a=215$ nm \cite{Lacroute2012}. $U_{\text{trap}}$ for the substates of the ground level $F=4$ of $6S_{1/2}$ (excited level $F'=5$ of $6P_{3/2}$) are shown as black (red-dashed) curves. \textbf{(a)(i)} azimuthal $U_{\text{trap}}(\phi)$, \textbf{(ii)} axial $U_{\text{trap}}(z)$ and \textbf{(b)} radial $U_{\text{trap}}(r-a)$ trapping potentials. Input polarizations for the trapping beams are denoted by the red and blue arrows in the inset in (b).}
%\noindent \hrulefill
\label{Utrap}%
\end{figure}

Following the landmark realization of a nanofiber trap \cite{Vetsch2010,Vetsch2010a,Dawkins2011}, in this Letter we report the implementation of a state-insensitive, compensated nanofiber trap for atomic Cesium (Cs), as illustrated in Fig. \ref{Utrap} \cite{Lacroute2012}. For our trap, differential scalar shifts $\delta U_{\text{scalar}}$ between ground and excited states are eliminated by using ``magic'' wavelengths for both red- and blue-detuned trapping fields \cite{ye08}. Inhomogeneous Zeeman broadening due to vector light shifts $\delta U_{\text{vector}}$ is suppressed by way of pairs of counter-propagating red- and blue-detuned fields.

\begin{figure}[tH!]
%\vspace{-4mm}
\centering
\includegraphics[width=1.\columnwidth]{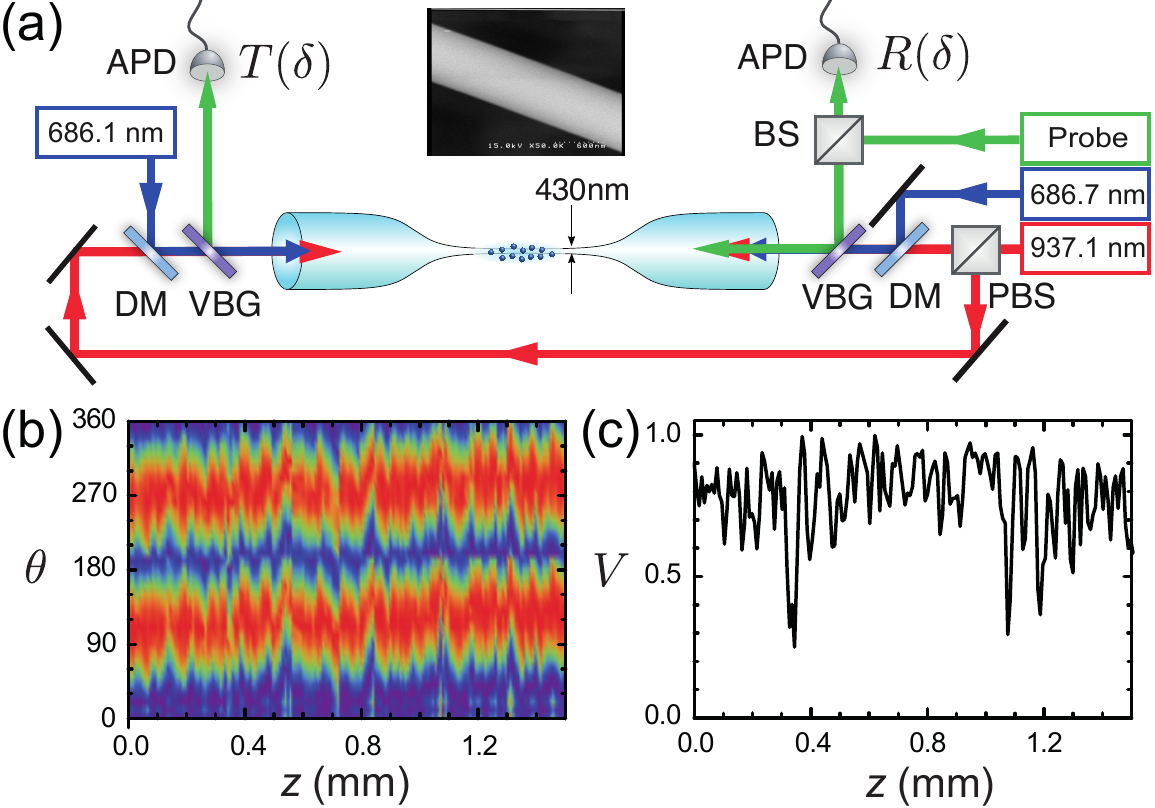}
\caption{Schematic of the setup for a state-insensitive, compensated nanofiber trap. VBG: Volume Bragg Grating, DM: Dichroic mirror, PBS: Polarizing beamsplitter, and APD: Avalanche photodetector. The inset shows a SEM image of the nanofiber for atom trapping. {\textbf {(b)}} Integrated intensity of the Rayleigh scattering from the nanofiber as a function of the angle $\theta$ of the polarization for the incident probe field and of distance $z$ along the fiber axis. {\textbf {(c)}} Spatially resolved visibility $V$ as a function of axial position $z$.}
%\noindent \hrulefill
\label{expsetup}
\end{figure}

The compensation of scalar and vector shifts results in a measured transition linewidth$\Gamma/2\pi = 5.7 \pm 0.1$ MHz for Cs atoms trapped $r_{\rm min}\simeq 215$ nm from the surface of an SiO$_2$ fiber of diameter $430$ nm, which should be compared to the free-space linewidth $\Gamma_0/2\pi = 5.2$ MHz for the $6S_{1/2},F=4$ $\rightarrow$ $6P_{3/2},F'=5$ Cs transition.  Probe light transmitted through the $1$D array of trapped atoms exhibits an optical depth $d_N=66\pm17$. From the measurements of optical depth and number $N$ of atoms, we infer a single-atom attenuation $d_1=d_N/N \simeq 0.08$. The bandwidth $\Gamma_R$ for reflection from the $1$D array is observed to broaden with increasing $N$, in direct proportion to the entropy for the multiplicity of trapping sites.

Our trapping scheme is based upon the analyses in Refs. \cite{Balykin2004,Kien2004a,Kien2005a,Nayak2007} and the demonstrations in Refs. \cite{Vetsch2010,Vetsch2010a,Dawkins2011}. A dielectric fiber in vacuum with radius $a$ smaller than a wavelength supports the ``hybrid'' fundamental mode $\text{HE}_{11}$, which carries significant energy in its evanescent field \cite{Tong2004}. For linear input polarization at angle $\phi_0$, an appropriate combination of attractive and repulsive $\text{HE}_{11}$ fields creates a dipole-force trap external to the fiber's surface with trap minima at $\phi-\phi_0=0,\pi$.

Following these principles, we have designed a ``magic compensation'' scheme that traps both ground and excited states and  greatly reduces the inhomogeneous broadening for atomic sublevels \cite{Lacroute2012}. By contrast, uncompensated schemes do not provide a stable trapping potential for excited states and suffer large dephasing between ground states over a single vibrational period \cite{Vetsch2010,Vetsch2010a,Dawkins2011}.

As shown in Fig. \ref{Utrap}, our trap consists of a pair of counter-propagating $x$-polarized red-detuned beams at the ``magic'' wavelength \cite{ye08} $\lambda_\text{red}=937$ nm to form an attractive $1$D optical lattice. A second pair of beams at the ``magic'' wavelength $\lambda_\text{blue}=686$ nm with detuning $\delta_\text{blue}$ provides a repulsive contribution to $U_{\text{trap}}$, thereby protecting the trapped atoms from the short-ranged attractive surface interaction. To avoid a standing wave incommensurate with that at $937$ nm, the blue-detuned beams have a relative detuning $\delta_{fb}=382$ GHz and effectively yield linearly polarized light at all positions; vector light shifts are suppressed by $\delta_{fb}/\delta_\text{blue} \simeq 4\times10^{-3}$ \cite{Lacroute2012}. The resulting potential $U_{\text{trap}}$ allows for state-insensitive, $3$D confinement of Cs atoms along a SiO$_2$ nanofiber for the $6S_{1/2}$ ground and $6P_{3/2}$ excited states.

We calculate the adiabatic potential $U_{\text{trap}}$ in Fig. \ref{Utrap} from the full light-shift Hamiltonian (i.e., scalar, vector, and tensor shifts), together with the surface potential for Casimir-Polder interactions with the dielectric \cite{Lacroute2012}. The red-detuned beams each have power $0.4$ mW, while the blue-detuned beams each have power $5$ mW. The trap depth at the minimum is $U_{\text{trap}}=-0.27$ mK located about $215$ nm from the fiber surface, with trap frequencies $\{w_{\rho},w_{z},w_{\phi}\}/2\pi=\{$199, 273, 35$\}$ kHz.

An overview of our experiment is given in Fig. \ref{expsetup}. A cloud of cold Cesium atoms (diameter $\sim 1$ mm) spatially overlaps the nanofiber. Cold atoms are loaded into $U_{\text{trap}}$ during an optical molasses phase  ($\sim 10$ ms) and are then optically pumped to $6S_{1/2},F=4$ for $0.5$ ms. The red- and blue-detuned trapping fields are constantly `on' throughout the laser cooling and loading processes with parameters comparable to those in Fig. \ref{Utrap}. 

For the transmission and reflection measurements, the trapped atoms are interrogated by a probe pulse ($1$ ms) with frequency $\omega_{P}$, optical power $P_{\text{probe}} \simeq 0.1$ pW and detuning $\delta=\omega_{P}-\omega_{A}$ relative to the $F=4\leftrightarrow F^{\prime}=5$ transition frequency $\omega_{A}$. The probe pulse is combined with the forward propagating trapping fields by a pair of Volume Bragg Gratings (VBGs) at the fiber input. The strong trapping beams are then filtered by a pair of VBGs at the fiber output, with the transmitted probe pulse monitored by a single-photon avalanche photodiode. The polarization of the probe laser is aligned along the trapping beams in order to maximize the overlap with the trapped atoms. We then shelve the atoms to $F=3$ with a depumping pulse, and probe the fiber transmission with a reference pulse to determine the input power of the probe pulse.

As described in the Appendix A, the polarizations of the trapping and probe beams are aligned by observations of Rayleigh scattering \cite{Vetsch2010}. From a simple model of the results in Fig. \ref{expsetup} (b, c), we infer a transverse polarization vector $\vec{E}_{\text{in}}(z=0)=(E_x,i E_y)$ with $\beta=\arctan(E_y/E_x) \simeq 12\pm3^\circ$ for the probe beam. The principal axes of the polarization ellipse rotate in an approximately linear fashion along $z$ in the trapping region by an angle $\phi(z) \simeq \phi_0+ (d\phi(z)/dz)\delta z$, where $\phi_0 \simeq 16^{\circ}$ and $d\phi(z)/dz \simeq 12^{\circ}/$mm. These results are incorporated into our analysis of the measured transmission and reflection spectra of the trapped atoms (Appendix B).

The lifetime of the atoms in our nanofiber trap is determined from the decay of the resonant optical depth $d_N\simeq N \sigma_0/A_{\text{eff}}$ as a function of storage time $\tau$. Here, $N$ is the number of trapped atoms, $\sigma_0$ is the resonant absorption cross-section, and $A_{\text{eff}}=P_{\text{probe}}/I_{\text{probe}}(\vec{r}_{\rm min})$ is the effective optical mode area of the probe's evanescent wave. We observe that $d_N$ decays exponentially with time constant $\tau_0=12\pm1$ ms. With pulsed polarization-gradient cooling, the lifetime is extended to $\tau_{\text{PG}}=140\pm11$ ms. We are currently characterizing the intensity and polarization noise spectra of the trapping fields to reduce parametric heating \cite{Savard1997} with the goal of extending the trap lifetime towards the limit $\tau_r \sim 30$ s set by recoil heating.

\begin{figure}[t!]
%\vspace{-4mm}
\centering
\includegraphics[width=0.95\columnwidth]{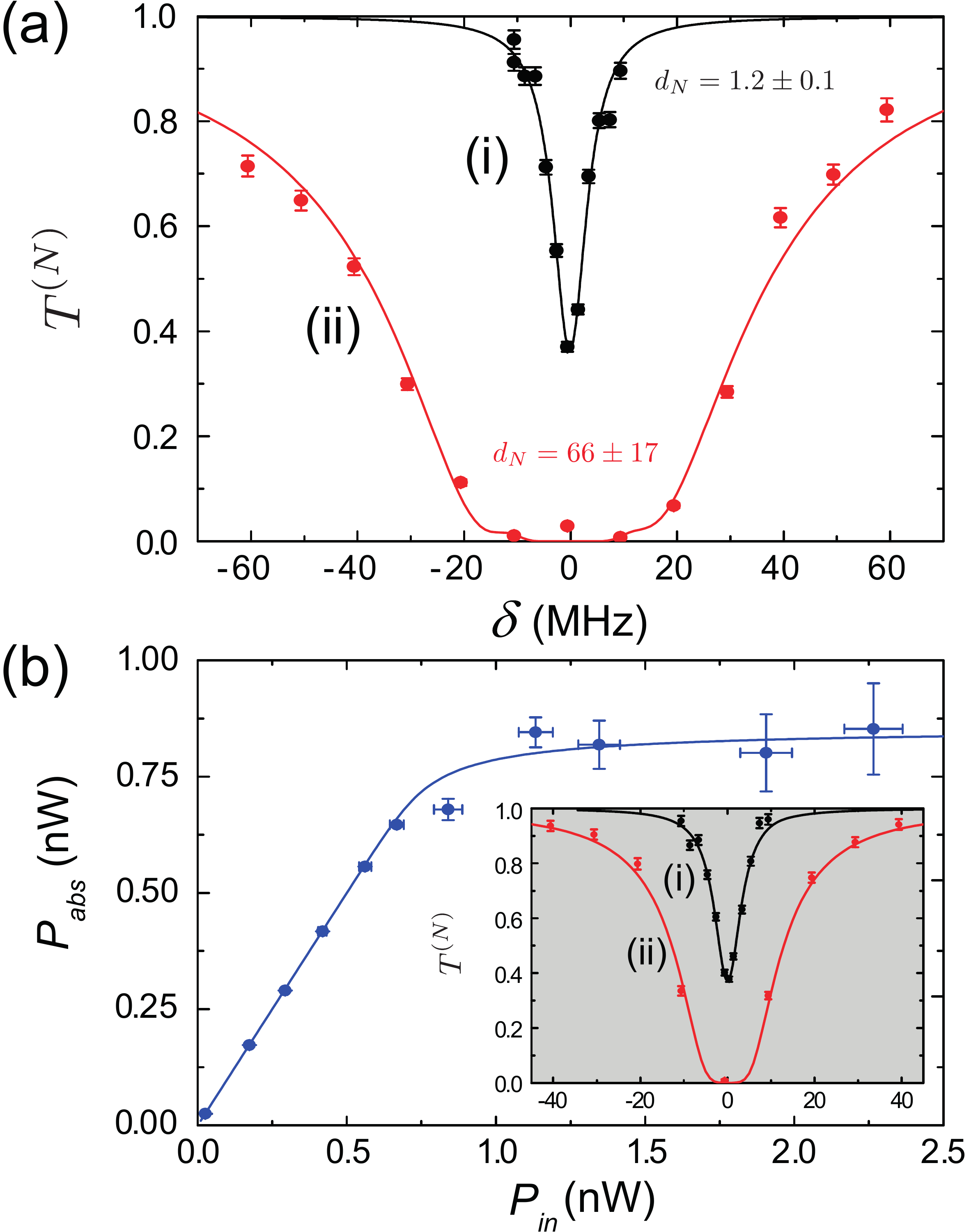}
\caption{\textbf{(a)} Probe transmission spectra $T^{(N)}(\delta)$ for $N$ trapped atoms as a function of detuning $\delta$ from the $6S_{1/2},F=4 \rightarrow 6P_{3/2}, F'=5$ transition in Cs for $x$-polarized input. From fits to $T^{(N)}(\delta)$ (full curves), we obtain the  optical depths $d_N$ at $\delta=0$ and linewidths $\Gamma$ from a model that incorporates the polarization measurements in Fig.\ref{expsetup} (b),(c) (Appendix B). $T^{(N)}(\delta)$ (i) at $\tau=299$ ms with $d_N=1.2\pm0.1$ and $\Gamma=5.8\pm0.5$ MHz and (ii) at $\tau=1$ ms with $d_N=66\pm17$. \textbf{(b)} Measurement of the power $P_{\text{abs}}$ absorbed by the trapped atoms as a function of input power $P_{\text{in}}$, together with the associated optical depth $d_N=18\pm2$ from curve (ii) (red line) of the inset, allow an inference of $N=224 \pm 10$ \cite{Vetsch2010}. The linewidth $\Gamma=5.5\pm0.4$ MHz and $d_N=1.1\pm0.1$ are determined from curve (i) (black line) of the inset.}
%\noindent \hrulefill
\label{transmissionfigure}%
\end{figure}

Fig. \ref{transmissionfigure}(a) displays the transmission spectra $T^{(N)}(\delta)$ for the $1$D atomic array. The linewidth $\Gamma=5.8\pm0.5$ MHz is determined from a model (solid black line)  incorporating fiber birefringence and  linear atomic susceptibility (Appendix B) in the low density regime ($\tau=299$ ms). The fitted line profile (solid red line) yields a maximum resonant optical depth $d_N=66\pm17$ at $\tau=1$ ms (Appendix B). Significantly, our magic, compensated scheme has no discernible shift of the transition frequency $\Delta_{A}/2\pi \simeq 0 \pm 0.5$ MHz relative to the free-space line center. The measured linewidths from $4$ data sets average to $\Gamma/2\pi \simeq 5.7 \pm 0.1$ MHz, as compared to the free-space radiative linewidth $\Gamma_0/2\pi=5.2$ MHz for the $6S_{1/2}\leftrightarrow 6P_{3/2}$ transition. By contrast, for a non-compensated scheme without magic wavelengths for Cs \cite{Vetsch2010}, the transition frequency is shifted by $\Delta_{A}/2\pi \simeq 13$ MHz and the linewidth is broadened to $\Gamma/2\pi \simeq 20$ MHz.

The broadening of the absorption linewidth above $\Gamma_0$ is predicted for our nanofiber trap because of the enhanced atomic decay into the forward and backward modes of the nanofiber at rate $\Gamma_{\text{1D}}$ \cite{Kien2005b}. We estimate that an atom at the minimum of $U_{\text{trap}}$ decays into the fiber at rate $\Gamma_{\text{1D}}^{\text{(th)}}/2\pi \simeq 0.35$ MHz, leading to a predicted linewidth $\Gamma_{\text{tot}}/2\pi \simeq 5.3$ MHz. Additional broadening arises from the tensor shifts of the excited $6P_{3/2}, F'=5$ manifold ($\simeq 0.7$ MHz) and Casimir-Polder shifts ($\simeq 0.1$ MHz), as well as technical noise of the probe laser ($\simeq 0.3$ MHz). While each of these contributions is being investigated in more detail, the spectra in Fig. \ref{transmissionfigure} provide strong support for the effectiveness of our state-insensitive, compensated trapping scheme \cite{Lacroute2012}.

To determine the number of trapped atoms, we carry out a saturation measurement at a storage time $\tau=1$ ms with $\delta=0$ MHz. As shown in Fig. \ref{transmissionfigure}(b), we measure the power $P_{\text{abs}}$ absorbed by the trapped atomic ensemble in the limit of high saturation $s=P/P_{\text{sat}}\gg 1$ \cite{Vetsch2010}. As described in the Appendix C, the fitted curve (blue solid line) yields a number of trapped atoms $N=224\pm10$. Together with the optical depth $d_N = 18$, we find an optical depth per atom $d_1=7.8\pm 1.3\%$ for Fig. \ref{transmissionfigure}(b) \cite{d1}. A similar measurement presented in the Appendix C with $d_N=43 \pm10$ and $N=564 \pm92$ yields $d_1=7.7\pm 2.2\%$. These measurements of $d_1$ and $\Gamma_{tot}$ were separated by four months and yield consistent results for the nanofiber trap. We thereby estimate $\Gamma_{\text{1D}}/2\pi \simeq 0.2$ MHz \cite{Chang2012}, similar to $\Gamma_{\text{1D}}^{\text{(th)}}$.

\begin{figure}[t!]
\centering
\includegraphics[width=0.95\columnwidth]{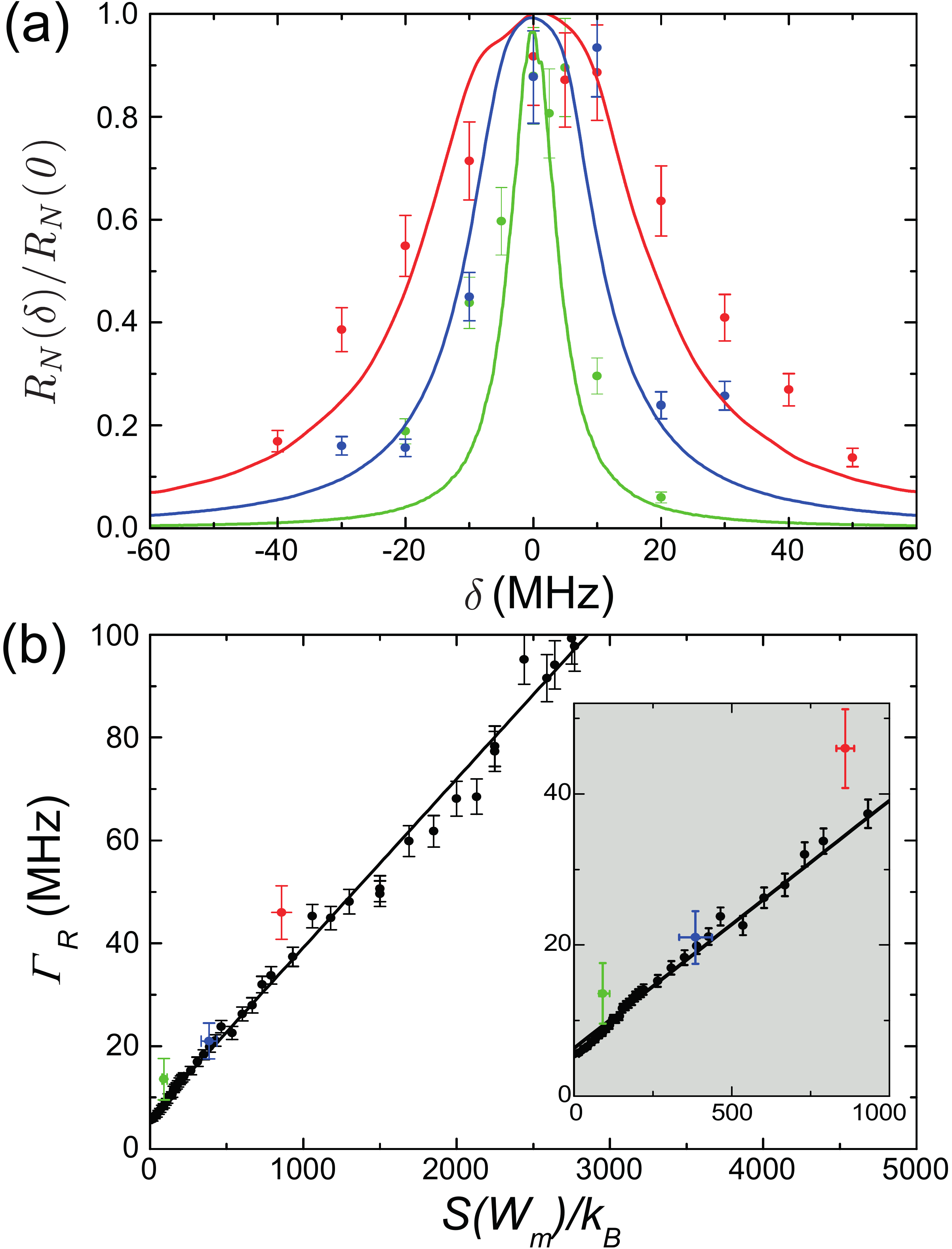}
\caption{\textbf{(a)} Normalized reflection spectra $R_N(\delta)/R_N(0)$ from the $1$D atomic array with $N=14\pm2$ atoms (green points), $N=79\pm13$ atoms (blue points) and $N=224\pm 10$ atoms (red points) randomly distributed across $n_{\text{site}}\simeq4000$ sites. The solid lines are the spectra obtained from Monte Carlo simulations for $N\ll n_{\text{site}}$. \textbf{(b)} Simulated linewidth $\Gamma_R$ of the reflection spectrum as a function of entropy $\mathcal{S}$. The error bars for the red, blue and green data points include both statistical and systematic uncertainties. The black points are the results of a simulation with error bars representing numerical uncertainties. The solid line is a linear fit to the simulation points.}
\label{reflectionfigure}%
\end{figure}

The reflection from the $1$D atomic array results from backscattering of the electromagnetic field within the $1$D system \cite{Chang2012}. The randomness in the distribution of $N$ atoms among $n_{\text{sites}}$ trapping sites can thus greatly affect the reflection spectrum $R_N(\delta)$. In Fig. \ref{reflectionfigure}(a), we observe $R_N(\delta)$ from the $1$D atomic array, where the measured Lorentzian linewidth $\Gamma_R$ is significantly broadened from $\Gamma_0$ for large $N$ (with $N\ll n_{\text{site}}$). 
The solid curves are for $R_N(\delta)$ from Monte-Carlo simulations for the atomic distribution based on the transfer matrix formalism \cite{Deutsch1995}.

In order to quantify the microscopic state of disorder for the system, we define an entropy $\mathcal{S}$ for the $1$D atomic array by $\mathcal{S}=\ln W_m(\psi)$, where  $W_m(\psi)=\frac{n_{\text{site}}!}{N !(n_{\text{site}}-N)!}$ is the multiplicity for the atomic distribution in the $1$D lattice. In Fig. \ref{reflectionfigure}(b), we find that the measured reflection linewidth $\Gamma_R$ (colored points) as well as the linewidth $\Gamma_R^{\text{th}}$ (black points) from the numerical simulation are proportional to the entropy $\mathcal{S}$ of the site-population statistics. These results demonstrate the strong modification of $R_N(\delta)$ due to randomness in the atomic distribution. 

In conclusion, we have trapped atoms along a nano-fiber by using a state-insensitive, compensated optical trap \cite{Lacroute2012} to achieve an optical depth $d_N \simeq 66$. Compared to previous work with hollow-core and nano-fibers, the atoms are trapped with small perturbations to dipole-allowed transitions. Our scheme is thus well-suited to various applications, including the creation of $1$D atomic mirrors for cavity QED and investigations of single-photon nonlinearities and quantum many-body physics in $1$D spin chains \cite{Chang2012}, as well as precision measurements of Casimir-Polder forces near a dielectric waveguide \cite{obrecht07}.

Currently, the maximum filling factor for sites over the $1$ mm loading region is $\sim 19\%$, which can be improved with adiabatic loading and elimination of collisional blockade \cite{Grunzweig2010}. The vibrational ground state for axial motion in $U_{\text{trap}}$ can be reached by introducing Raman sidebands on the $937$ nm trapping fields \cite{Boozer06}. The strong axial confinement in our trap implies the presence of a large anharmonicity in the vibrational ladder, which could provide a tool for experiments with single phonons. Furthermore, the design principles of our magic, compensated trap can be extended from simple `nanowires' to complex photonic crystal structures \cite{eichenfield_optomechanical_2009}.

We gratefully acknowledge the contributions of D. Chang, I. Cirac, A. Gorshkov, L. Jiang, J. A. Muniz Silva, and E. Polzik.
 Funding at Caltech is provided by the IQIM, an NSF Physics Frontier Center with support of the Gordon and Betty Moore Foundation, by the AFOSR QuMPASS MURI, by the DoD NSSEFF program, and by NSF Grant $\#$ PHY0652914. AG is supported by the Nakajima Foundation. The research of KSC at KIST is supported by the KIST institutional program.

\appendix
\section{Fiber  characterization}

\vspace{-3mm}
\subsection{Fiber diameter measurement}
\vspace{-3mm}

\begin{figure}[b!]
\centering
\includegraphics[width=1.0\columnwidth]{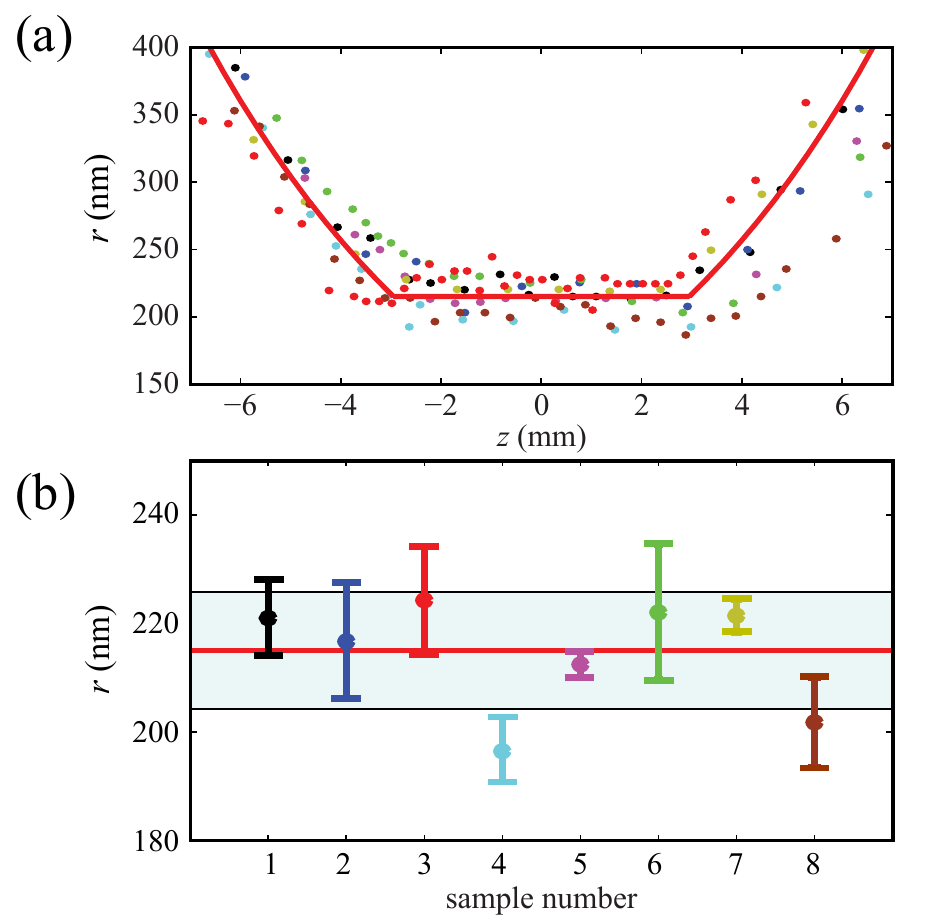}
\caption{ \textbf{(a)} The radius $r(z)$ as a function of axial coordinate $z$ from SEM measurements for a set of $8$ fiber tapers. \textbf{(b)} Radius $r(z)$ for the central region, which yields an average radius $r(z)=215\pm10$ nm for $-3<z<3$ mm.}
\label{taperfigure}%
\end{figure}

The tapered SiO$_2$ nanofiber for the trap is drawn from a conventional optical fiber in a hydrogen-oxygen flame \cite{Birks1992,Stiebeiner2010,Aoki2010}. 
Fig. \ref{taperfigure} shows the radius $r(z)$ as a function of axial coordinate $z$ from SEM measurements for a set of $8$ fiber tapers. The central region $\Delta z$ yields an average radius $r(z)=215\pm10$ nm for $-3<z<3$ mm, as shown in (b). The red curve is a theoretical calculation for $r(z)$ \cite{Birks1992} with the relevant taper pulling parameters and an effective heating length $l_H=5.9$ mm.
%{ of the heated region for drawing the tapered fiber as the only adjustable parameter.}
Independent measurement gives $l_H =6.1\pm0.9$ mm.

\vspace{-5mm}
\subsection{Polarization measurement}
\begin{figure}[t!]
\centering
\includegraphics[width=1.0\columnwidth]{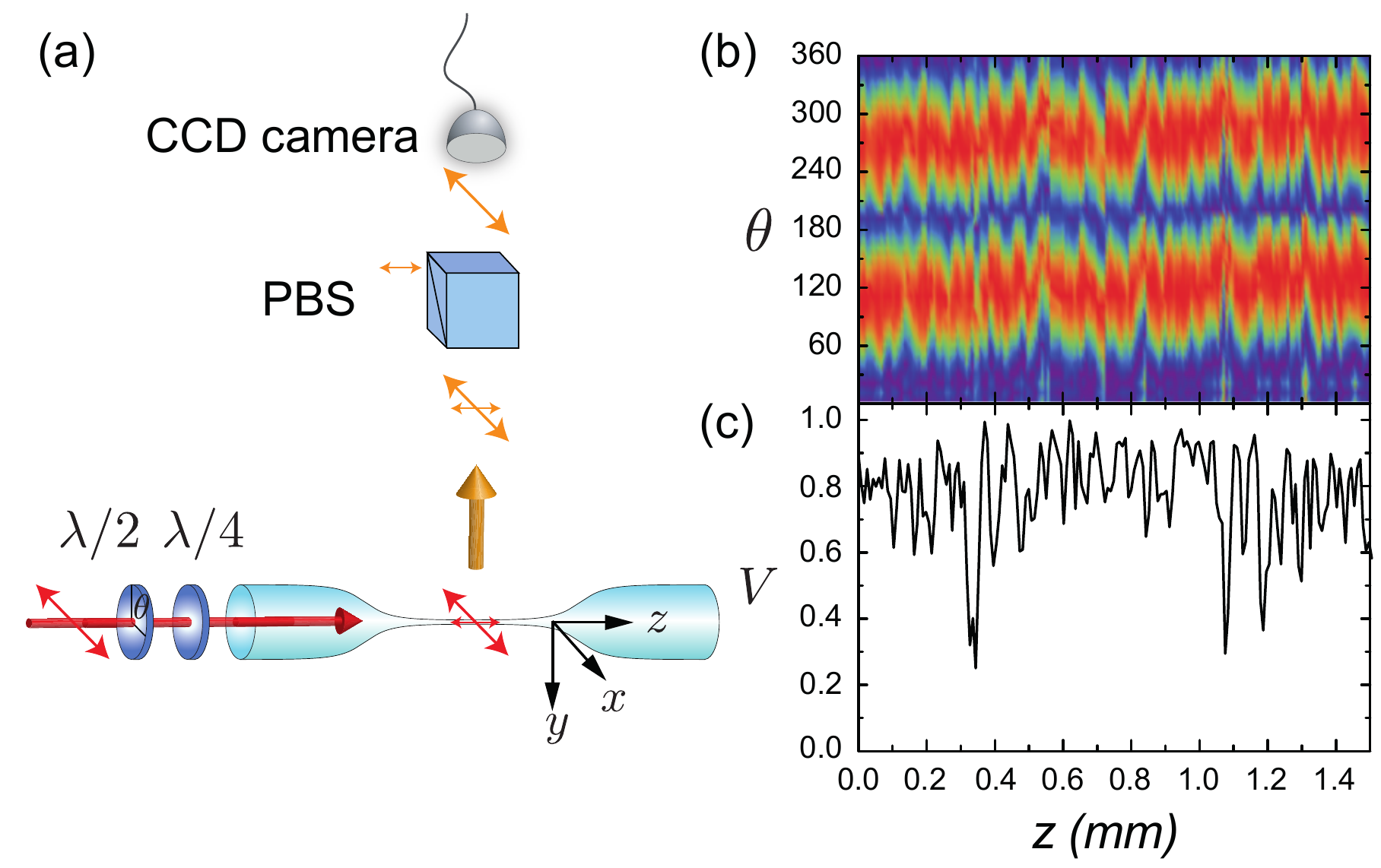}
\caption{\textbf{(a)} Setup for polarization characterization of the fields along the nanofiber. \textbf {(b)} Integrated intensity $I(z,\theta)$ of scattering from the nanofiber as a function of the angle $\theta$ of a half-waveplate for the incident probe field and of distance $z$ along the fiber axis. {\textbf {(c)}} Spatially resolved visibility $V$ as a function of axial position $z$.}
\label{polarizationfigure}%
\end{figure}

\vspace{-3mm}
The set-up and procedure for measuring and optimizing the polarization state of the nanofiber-guided fields are illustrated
in Fig. \ref{polarizationfigure} \cite{Vetsch2010a,Dawkins2011}. The angular distribution of the Rayleigh scattered light emitted from the nanofiber is observed
by a CCD camera aligned perpendicular to the fiber axis. For the radiation pattern $I(\varphi)\approx\sin^2(\varphi)$ of a dipole induced by linearly polarized light, no light should be detected along the direction that the dipoles oscillate ($\varphi= 0, \pi$). Since the HE$_{11}$ mode has a non-transversal polarization along the fiber axis $z$, a polarizer (PBS) is placed in front of a CCD camera to block  $z$-polarized light. By rotating the angle $\theta$ for the input polarization of the probe field with the half-waveplate ($\lambda/2$) shown in Fig. \ref{polarizationfigure}(a), a sinusoidal modulation of the scattering intensities $I(z,\theta)$ is observed along the $z$ axis of the fiber. From $I(z,\theta)$, we determine the visibility $V(z)=\frac{I(z,\theta)_{\text{max}}-I(z,\theta)_{\text{min}}}{I(z,\theta)_{\text{max}}+I(z,\theta)_{\text{min}}}$ as a function of $z$ along the nanofiber as shown in in Fig. \ref{polarizationfigure}(c).

To describe these measurements, we use a simple model based upon $z$-dependent birefringence in the fiber that results in elliptical polarization for the transverse fields $E_{x,y}$ along $z$. From Fig. \ref{polarizationfigure}(b) in the region of the atom trap, we deduce that the principal axes of the polarization ellipse rotate along $z$ by an angle $\phi(z) \simeq \phi_0+ (d\phi(z)/dz)\delta z$, where $\phi_0$ describes a possible offset from the symmetry axes of the trap (i.e., the direction 
$\phi=0$ in Fig. 1) and $(d\phi(z)/dz)=12^\circ/$mm gives the variation from this offset along $z$. From fits to transmission spectra described in the next section, we find that $\phi_0 \simeq 16^\circ$.

Fig. \ref{polarizationfigure}(c) makes clear that there is a wide range of visibility values $V(z)$ for the probe beam along $z$. For our simple model, we determine the ellipticity of the transverse polarization vector $\vec{E}_{x,y}(z)=(E_x(z),i E_y(z))$, which we describe by the phase $\beta(z)=\arctan(E_y(z)/E_x(z))$. Absent trapped atoms, the principal axes for the polarization ellipse also rotate by angle $\phi(z)$ along the nanofiber. From fits to transmission spectra, we find that fixed $\beta(z)=\beta_0$ provides an adequate description of our measurements, with $\beta_0 \simeq 12\pm3^\circ$ determined along the high-visibility region in Fig. \ref{polarizationfigure}(c).

\vspace{-4mm}
\section{Optical depth estimation}
\vspace{-3mm}

To model the combined effects of the linear atomic susceptibility (i.e., absorption and dispersion) together with the spatially varying birefringence along the nanofiber, we divide the region $0\leqslant z \leqslant L$ into $M$ `cells' each of length $\delta z=L/M$. Propagation through a cell at $z_j-\delta z/2 \leqslant z_j \leqslant z_j + \delta z/2$ is generated by the product of two transfer matrices: (1) $\hat{R}(\phi(z))$ produces rotation of the polarization axes from $\phi(z_j)$ to $\phi(z_j) + \delta \phi$, with $\delta \phi = (d\phi(z)/dz)\delta z$, and (2) $T(\delta,d_M)$ specifies the propagation of $E_x(z_j),E_y(z_j)$ through the trapped atoms with detuning $\delta$ and optical depth $d_M$ for the $x,y$ polarizations. Here $d_M=d_N/M$ is the optical depth of a single cell, and $d_N$ is the optical depth for the entire sample of $N$ atoms. For each cell, we take the ratio of resonant optical depths to be $d_{0, x}/d_{0, y}=2.5$ at the trap minimum $r_{\text{min}}=215$ nm, given by the asymmetry of the evanescent HE$_{11}$ fiber mode \cite{Dawkins2011}, where $x,y$ refer to the coordinates for the trap axis as in Fig. 1.

\begin{figure}[t!]
\centering
\includegraphics[width=1.0\columnwidth]{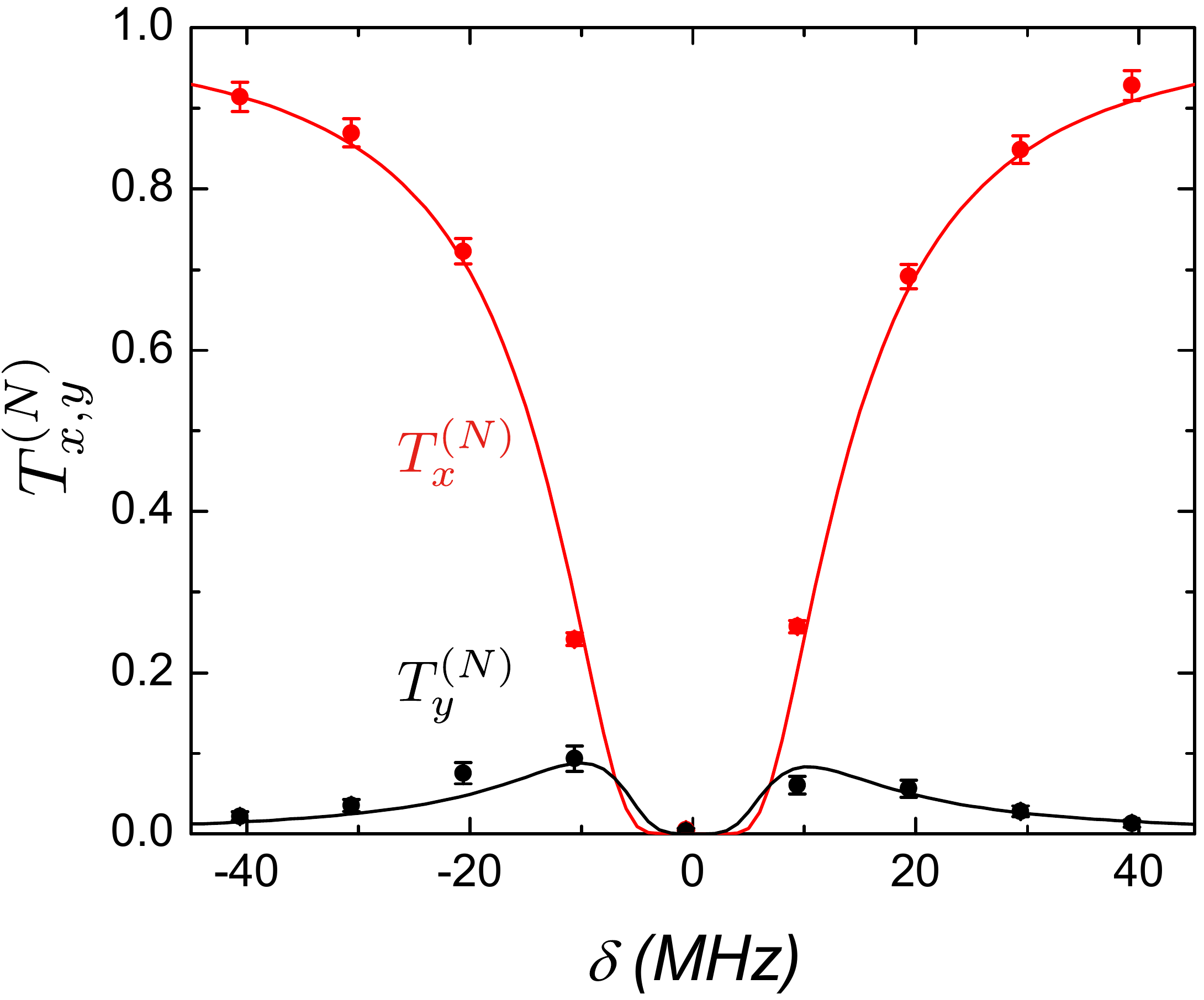}
\caption{Measured transmission spectra $T^{(N)}_{x}$ (red points) and  $T^{(N)}_y$(black points) for $x,y$ polarizations of the probe beam as functions of the detuning $\delta$ from the $|F=4\rangle\leftrightarrow|F'=5\rangle$ transition. From theoretical fits of our model (full curves) that include the interplay of spatially varying birefringence (as in Fig. \ref{polarizationfigure}) and of atomic absorption and dispersion along the nanofiber, we infer an optical depth $d_N=18 \pm 2$ with polarization offset $\phi_0=16^{\circ} \pm 3^{\circ}$ and
$\beta_0=12^\circ$.}
\label{polrotfig}
\end{figure}

In correspondence to the results from Fig. \ref{polarizationfigure}(b, c), the input polarization to this sequence of cells is set to be elliptically polarized with angle $\beta_0$ and offset $\phi_0$. After propagation through the fiber in the absence of atoms, our experimental procedure is to compensate the state of the output polarization with the transformation matrix $\hat{R}_{\text{out}}$ to maximize the polarization contrast for the photo-detection after a polarization beamsplitter aligned along $x,y$. Hence, in our model, a final rotation $\hat{R}_{\text{out}}$ follows propagation through the $M$ cells before projection of the output fields to obtain intensities $I_x^{\text{out}},I_y^{\text{out}}$.

Figure \ref{polrotfig} displays measured transmission spectra $T_{x,y}^{(N)}$ together with fits to the spectra by way of our model. The input parameters for the theoretical spectra are the polarization angle $\beta_0$, offset $\phi_0$, and optical depth $d_N$. Typically, we fix $\beta_0=12^\circ$ from our measurements of visibility and perform a least-square minimization over $\phi_0$ and $d_N$.

\begin{figure}[t!]
\centering
\includegraphics[width=1.0\columnwidth]{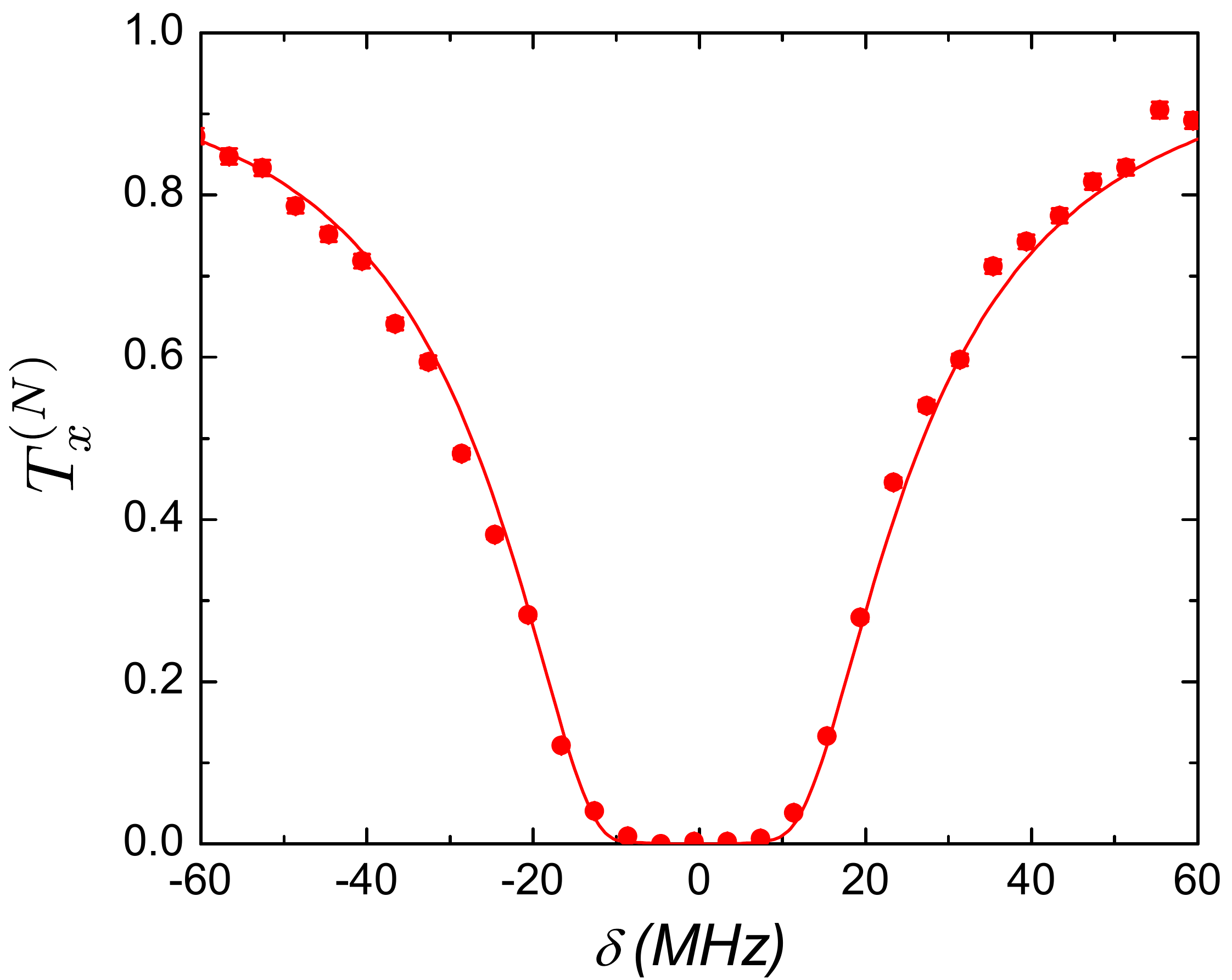}
\caption{Measured transmission spectrum $T^{(N)}_{x}$ (red points) for $x$ polarization of the probe beam as a function of the detuning $\delta$ from the $|F=4\rangle\leftrightarrow|F'=5\rangle$ transition. From a theoretical fit of our model (full curve) that includes spatially varying birefringence (as in Fig. \ref{polarizationfigure}) and atomic absorption and dispersion along the nanofiber, we infer an optical depth $d_N=43 \pm 10$ with polarization offset $\phi_0=16^{\circ} \pm 9^{\circ}$ and $\beta_0=12^\circ $.}
%\noindent \hrulefill
\label{transfig}
\end{figure}

From fits to the data in Fig. \ref{polrotfig}, we find $d_N=18 \pm 2$ and $\phi_0=16^{\circ} \pm 3^{\circ}$ for $\beta_0 = 12^{\circ}$. The quoted uncertainty for $d_N$ includes the contributions from the statistical uncertainty from the photoelectric detection statistics. A second example comparing measured and theoretical transmission spectra is given in Figure \ref{transfig}, now for the $x$ polarization at high optical density. From fits to the data in Fig. \ref{transfig}, we find $d_N=43\pm 10$ and $\phi_0=16^{\circ} \pm 9^{\circ}$ for $\beta_0 = 12^{\circ}$.

The theoretical fits to the transmission spectra shown in Fig. 2 have been obtained by applying this  model. From these and other fits to the measured transmission spectra, we have consistently found that $\overline{\phi_0} \simeq 16^\circ \pm 2^\circ$ with fixed $\beta_0=12^\circ$. Numerically, we find that small variations in $\phi_0$ can be compensated by corresponding changes in $\beta_0$ around the optimal values found from the fits.

\vspace{-4mm}
\section{Estimating the number of trapped atoms}
\vspace{-3mm}

A generalized Beer's law \cite{Vetsch2010,Vetsch2010a},

\begin{equation}
\label{beers}
\frac{dP(z)}{dz}=-n_z\frac{\sigma_0}{A_{\text{eff}}}\frac{P(z)}{1+P(z)/P_{\text{sat}}},
\end{equation}
describes the saturation behavior of a trapped atom. Here, the saturation power is $P_{\text{sat}}=I_{\text{sat}}A_{\text{eff}}=49.6$ pW, effective optical mode area is $A_{\text{eff}}=1.8$ $\mu$m$^2$, $\sigma_0$ is the resonant absorption cross-section, and $n_z=N/L$ is the atomic line density for a sample length $L=1$ mm.

For measurements as in Fig. 3(a), the number of trapped atoms $N=224 \pm10$ is deduced by fitting the data with the solution of the generalized Beer's law, Eq. \ref{beers}. Together with the optical depth $d_N=18 \pm 2$ from the curve (ii) of Fig. 3(b), we infer an optical depth $d_1=7.8 \pm 1.3 \%$ for a single atom.

Similar measurements taken in conjunction with the transmission spectrum $T^{(N)}_{x}$ in Fig. \ref{transfig} lead to an estimate $N=564 \pm 92$ for the number of atoms trapped along the nanofiber. Together with the total optical depth $d_N=43 \pm 10$ from Fig. \ref{transfig}, we infer an optical depth $d_1=7.7 \pm 2.2\%$ for a single atom.

$^\ast$These authors contributed equally to this work.

$^{\dagger}$Current address: Department of Physics, ETH Z\"{u}rich, CH-8093 Z\"{u}rich, Switzerland.

$^\ddag$Current address: Department of Physics and Astronomy, Northwestern University, Evanston, IL 60208.

\end{document}